\documentclass[11pt]{article}

    \usepackage[breakable]{tcolorbox}
    \usepackage{parskip} 
    
    \usepackage{iftex}
    \ifPDFTeX
    	\usepackage[T1]{fontenc}
    	\usepackage{mathpazo}
    \else
    	\usepackage{fontspec}
    \fi

    \usepackage{graphicx}
    
    \usepackage{caption}
    \DeclareCaptionFormat{nocaption}{}
    \usepackage[Export]{adjustbox} 
    \adjustboxset{max size={0.9\linewidth}{0.9\paperheight}}
    \usepackage{float}
    \usepackage{xcolor} 
    \usepackage{enumerate} 
    \usepackage{geometry} 
    \usepackage{amsmath} 
    \usepackage{amssymb} 
    \usepackage{textcomp} 
    \AtBeginDocument{%
    }
    \usepackage{upquote} 
    \usepackage{eurosym} 
    \usepackage[mathletters]{ucs} 
    \usepackage{fancyvrb} 
    \usepackage{grffile} 
    \makeatletter 
    \def\Gread@@xetex#1{%
      \IfFileExists{"\Gin@base".bb}%
      {\Gread@eps{\Gin@base.bb}}%
      {\Gread@@xetex@aux#1}%
    }
    \makeatother

    \usepackage{hyperref}
    \usepackage{titling}
    \usepackage{longtable} 
    \usepackage{booktabs}  
    \usepackage[inline]{enumitem} 
    \usepackage[normalem]{ulem} 
    \usepackage{mathrsfs}

    \definecolor{urlcolor}{rgb}{0,.145,.698}
    \definecolor{linkcolor}{rgb}{.71,0.21,0.01}
    \definecolor{citecolor}{rgb}{.12,.54,.11}

    \definecolor{ansi-black}{HTML}{3E424D}
    \definecolor{ansi-black-intense}{HTML}{282C36}
    \definecolor{ansi-red}{HTML}{E75C58}
    \definecolor{ansi-red-intense}{HTML}{B22B31}
    \definecolor{ansi-green}{HTML}{00A250}
    \definecolor{ansi-green-intense}{HTML}{007427}
    \definecolor{ansi-yellow}{HTML}{DDB62B}
    \definecolor{ansi-yellow-intense}{HTML}{B27D12}
    \definecolor{ansi-blue}{HTML}{208FFB}
    \definecolor{ansi-blue-intense}{HTML}{0065CA}
    \definecolor{ansi-magenta}{HTML}{D160C4}
    \definecolor{ansi-magenta-intense}{HTML}{A03196}
    \definecolor{ansi-cyan}{HTML}{60C6C8}
    \definecolor{ansi-cyan-intense}{HTML}{258F8F}
    \definecolor{ansi-white}{HTML}{C5C1B4}
    \definecolor{ansi-white-intense}{HTML}{A1A6B2}
    \definecolor{ansi-default-inverse-fg}{HTML}{FFFFFF}
    \definecolor{ansi-default-inverse-bg}{HTML}{000000}

    \DefineVerbatimEnvironment{Highlighting}{Verbatim}{commandchars=\\\{\}}

    \let\Oldtex\TeX
    \let\Oldlatex\LaTeX
    \renewcommand{\TeX}{\textrm{\Oldtex}}
    \renewcommand{\LaTeX}{\textrm{\Oldlatex}}
    \title{Using \texttt{bayesmixedlogit} and \texttt{bayesmixedlogitwtp} in Stata}
    \author{Matthew J. Baker \\ Hunter College and the Graduate Center, CUNY}

\makeatletter
\def\PY@reset{\let\PY@it=\relax \let\PY@bf=\relax%
    \let\PY@ul=\relax \let\PY@tc=\relax%
    \let\PY@bc=\relax \let\PY@ff=\relax}
\def\PY@tok#1{\csname PY@tok@#1\endcsname}
\def\PY@toks#1+{\ifx\relax#1\empty\else%
    \PY@tok{#1}\expandafter\PY@toks\fi}
\def\PY@do#1{\PY@bc{\PY@tc{\PY@ul{%
    \PY@it{\PY@bf{\PY@ff{#1}}}}}}}
\def\PY#1#2{\PY@reset\PY@toks#1+\relax+\PY@do{#2}}

\expandafter\def\csname PY@tok@w\endcsname{\def\PY@tc##1{\textcolor[rgb]{0.73,0.73,0.73}{##1}}}
\expandafter\def\csname PY@tok@c\endcsname{\let\PY@it=\textit\def\PY@tc##1{\textcolor[rgb]{0.25,0.50,0.50}{##1}}}
\expandafter\def\csname PY@tok@cp\endcsname{\def\PY@tc##1{\textcolor[rgb]{0.74,0.48,0.00}{##1}}}
\expandafter\def\csname PY@tok@k\endcsname{\let\PY@bf=\textbf\def\PY@tc##1{\textcolor[rgb]{0.00,0.50,0.00}{##1}}}
\expandafter\def\csname PY@tok@kp\endcsname{\def\PY@tc##1{\textcolor[rgb]{0.00,0.50,0.00}{##1}}}
\expandafter\def\csname PY@tok@kt\endcsname{\def\PY@tc##1{\textcolor[rgb]{0.69,0.00,0.25}{##1}}}
\expandafter\def\csname PY@tok@o\endcsname{\def\PY@tc##1{\textcolor[rgb]{0.40,0.40,0.40}{##1}}}
\expandafter\def\csname PY@tok@ow\endcsname{\let\PY@bf=\textbf\def\PY@tc##1{\textcolor[rgb]{0.67,0.13,1.00}{##1}}}
\expandafter\def\csname PY@tok@nb\endcsname{\def\PY@tc##1{\textcolor[rgb]{0.00,0.50,0.00}{##1}}}
\expandafter\def\csname PY@tok@nf\endcsname{\def\PY@tc##1{\textcolor[rgb]{0.00,0.00,1.00}{##1}}}
\expandafter\def\csname PY@tok@nc\endcsname{\let\PY@bf=\textbf\def\PY@tc##1{\textcolor[rgb]{0.00,0.00,1.00}{##1}}}
\expandafter\def\csname PY@tok@nn\endcsname{\let\PY@bf=\textbf\def\PY@tc##1{\textcolor[rgb]{0.00,0.00,1.00}{##1}}}
\expandafter\def\csname PY@tok@ne\endcsname{\let\PY@bf=\textbf\def\PY@tc##1{\textcolor[rgb]{0.82,0.25,0.23}{##1}}}
\expandafter\def\csname PY@tok@nv\endcsname{\def\PY@tc##1{\textcolor[rgb]{0.10,0.09,0.49}{##1}}}
\expandafter\def\csname PY@tok@no\endcsname{\def\PY@tc##1{\textcolor[rgb]{0.53,0.00,0.00}{##1}}}
\expandafter\def\csname PY@tok@nl\endcsname{\def\PY@tc##1{\textcolor[rgb]{0.63,0.63,0.00}{##1}}}
\expandafter\def\csname PY@tok@ni\endcsname{\let\PY@bf=\textbf\def\PY@tc##1{\textcolor[rgb]{0.60,0.60,0.60}{##1}}}
\expandafter\def\csname PY@tok@na\endcsname{\def\PY@tc##1{\textcolor[rgb]{0.49,0.56,0.16}{##1}}}
\expandafter\def\csname PY@tok@nt\endcsname{\let\PY@bf=\textbf\def\PY@tc##1{\textcolor[rgb]{0.00,0.50,0.00}{##1}}}
\expandafter\def\csname PY@tok@nd\endcsname{\def\PY@tc##1{\textcolor[rgb]{0.67,0.13,1.00}{##1}}}
\expandafter\def\csname PY@tok@s\endcsname{\def\PY@tc##1{\textcolor[rgb]{0.73,0.13,0.13}{##1}}}
\expandafter\def\csname PY@tok@sd\endcsname{\let\PY@it=\textit\def\PY@tc##1{\textcolor[rgb]{0.73,0.13,0.13}{##1}}}
\expandafter\def\csname PY@tok@si\endcsname{\let\PY@bf=\textbf\def\PY@tc##1{\textcolor[rgb]{0.73,0.40,0.53}{##1}}}
\expandafter\def\csname PY@tok@se\endcsname{\let\PY@bf=\textbf\def\PY@tc##1{\textcolor[rgb]{0.73,0.40,0.13}{##1}}}
\expandafter\def\csname PY@tok@sr\endcsname{\def\PY@tc##1{\textcolor[rgb]{0.73,0.40,0.53}{##1}}}
\expandafter\def\csname PY@tok@ss\endcsname{\def\PY@tc##1{\textcolor[rgb]{0.10,0.09,0.49}{##1}}}
\expandafter\def\csname PY@tok@sx\endcsname{\def\PY@tc##1{\textcolor[rgb]{0.00,0.50,0.00}{##1}}}
\expandafter\def\csname PY@tok@m\endcsname{\def\PY@tc##1{\textcolor[rgb]{0.40,0.40,0.40}{##1}}}
\expandafter\def\csname PY@tok@gh\endcsname{\let\PY@bf=\textbf\def\PY@tc##1{\textcolor[rgb]{0.00,0.00,0.50}{##1}}}
\expandafter\def\csname PY@tok@gu\endcsname{\let\PY@bf=\textbf\def\PY@tc##1{\textcolor[rgb]{0.50,0.00,0.50}{##1}}}
\expandafter\def\csname PY@tok@gd\endcsname{\def\PY@tc##1{\textcolor[rgb]{0.63,0.00,0.00}{##1}}}
\expandafter\def\csname PY@tok@gi\endcsname{\def\PY@tc##1{\textcolor[rgb]{0.00,0.63,0.00}{##1}}}
\expandafter\def\csname PY@tok@gr\endcsname{\def\PY@tc##1{\textcolor[rgb]{1.00,0.00,0.00}{##1}}}
\expandafter\def\csname PY@tok@ge\endcsname{\let\PY@it=\textit}
\expandafter\def\csname PY@tok@gs\endcsname{\let\PY@bf=\textbf}
\expandafter\def\csname PY@tok@gp\endcsname{\let\PY@bf=\textbf\def\PY@tc##1{\textcolor[rgb]{0.00,0.00,0.50}{##1}}}
\expandafter\def\csname PY@tok@go\endcsname{\def\PY@tc##1{\textcolor[rgb]{0.53,0.53,0.53}{##1}}}
\expandafter\def\csname PY@tok@gt\endcsname{\def\PY@tc##1{\textcolor[rgb]{0.00,0.27,0.87}{##1}}}
\expandafter\def\csname PY@tok@err\endcsname{\def\PY@bc##1{\setlength{\fboxsep}{0pt}\fcolorbox[rgb]{1.00,0.00,0.00}{1,1,1}{\strut ##1}}}
\expandafter\def\csname PY@tok@kc\endcsname{\let\PY@bf=\textbf\def\PY@tc##1{\textcolor[rgb]{0.00,0.50,0.00}{##1}}}
\expandafter\def\csname PY@tok@kd\endcsname{\let\PY@bf=\textbf\def\PY@tc##1{\textcolor[rgb]{0.00,0.50,0.00}{##1}}}
\expandafter\def\csname PY@tok@kn\endcsname{\let\PY@bf=\textbf\def\PY@tc##1{\textcolor[rgb]{0.00,0.50,0.00}{##1}}}
\expandafter\def\csname PY@tok@kr\endcsname{\let\PY@bf=\textbf\def\PY@tc##1{\textcolor[rgb]{0.00,0.50,0.00}{##1}}}
\expandafter\def\csname PY@tok@bp\endcsname{\def\PY@tc##1{\textcolor[rgb]{0.00,0.50,0.00}{##1}}}
\expandafter\def\csname PY@tok@fm\endcsname{\def\PY@tc##1{\textcolor[rgb]{0.00,0.00,1.00}{##1}}}
\expandafter\def\csname PY@tok@vc\endcsname{\def\PY@tc##1{\textcolor[rgb]{0.10,0.09,0.49}{##1}}}
\expandafter\def\csname PY@tok@vg\endcsname{\def\PY@tc##1{\textcolor[rgb]{0.10,0.09,0.49}{##1}}}
\expandafter\def\csname PY@tok@vi\endcsname{\def\PY@tc##1{\textcolor[rgb]{0.10,0.09,0.49}{##1}}}
\expandafter\def\csname PY@tok@vm\endcsname{\def\PY@tc##1{\textcolor[rgb]{0.10,0.09,0.49}{##1}}}
\expandafter\def\csname PY@tok@sa\endcsname{\def\PY@tc##1{\textcolor[rgb]{0.73,0.13,0.13}{##1}}}
\expandafter\def\csname PY@tok@sb\endcsname{\def\PY@tc##1{\textcolor[rgb]{0.73,0.13,0.13}{##1}}}
\expandafter\def\csname PY@tok@sc\endcsname{\def\PY@tc##1{\textcolor[rgb]{0.73,0.13,0.13}{##1}}}
\expandafter\def\csname PY@tok@dl\endcsname{\def\PY@tc##1{\textcolor[rgb]{0.73,0.13,0.13}{##1}}}
\expandafter\def\csname PY@tok@s2\endcsname{\def\PY@tc##1{\textcolor[rgb]{0.73,0.13,0.13}{##1}}}
\expandafter\def\csname PY@tok@sh\endcsname{\def\PY@tc##1{\textcolor[rgb]{0.73,0.13,0.13}{##1}}}
\expandafter\def\csname PY@tok@s1\endcsname{\def\PY@tc##1{\textcolor[rgb]{0.73,0.13,0.13}{##1}}}
\expandafter\def\csname PY@tok@mb\endcsname{\def\PY@tc##1{\textcolor[rgb]{0.40,0.40,0.40}{##1}}}
\expandafter\def\csname PY@tok@mf\endcsname{\def\PY@tc##1{\textcolor[rgb]{0.40,0.40,0.40}{##1}}}
\expandafter\def\csname PY@tok@mh\endcsname{\def\PY@tc##1{\textcolor[rgb]{0.40,0.40,0.40}{##1}}}
\expandafter\def\csname PY@tok@mi\endcsname{\def\PY@tc##1{\textcolor[rgb]{0.40,0.40,0.40}{##1}}}
\expandafter\def\csname PY@tok@il\endcsname{\def\PY@tc##1{\textcolor[rgb]{0.40,0.40,0.40}{##1}}}
\expandafter\def\csname PY@tok@mo\endcsname{\def\PY@tc##1{\textcolor[rgb]{0.40,0.40,0.40}{##1}}}
\expandafter\def\csname PY@tok@ch\endcsname{\let\PY@it=\textit\def\PY@tc##1{\textcolor[rgb]{0.25,0.50,0.50}{##1}}}
\expandafter\def\csname PY@tok@cm\endcsname{\let\PY@it=\textit\def\PY@tc##1{\textcolor[rgb]{0.25,0.50,0.50}{##1}}}
\expandafter\def\csname PY@tok@cpf\endcsname{\let\PY@it=\textit\def\PY@tc##1{\textcolor[rgb]{0.25,0.50,0.50}{##1}}}
\expandafter\def\csname PY@tok@c1\endcsname{\let\PY@it=\textit\def\PY@tc##1{\textcolor[rgb]{0.25,0.50,0.50}{##1}}}
\expandafter\def\csname PY@tok@cs\endcsname{\let\PY@it=\textit\def\PY@tc##1{\textcolor[rgb]{0.25,0.50,0.50}{##1}}}

\def\PYZbs{\char`\\}
\def\PYZus{\char`\_}

\makeatother

    \makeatletter
        \newbox\Wrappedcontinuationbox 
        \newbox\Wrappedvisiblespacebox 
        \newcommand*\Wrappedvisiblespace {\textcolor{red}{\textvisiblespace}} 
        \newcommand*\Wrappedcontinuationsymbol {\textcolor{red}{\llap{\tiny$\m@th\hookrightarrow$}}} 
        \newcommand*\Wrappedcontinuationindent {3ex } 
        \newcommand*\Wrappedafterbreak {\kern\Wrappedcontinuationindent\copy\Wrappedcontinuationbox} 
        \newcommand*\Wrappedbreaksatspecials {%
            \def\PYGZus{\discretionary{\char`\_}{\Wrappedafterbreak}{\char`\_}}%
            \def\PYGZob{\discretionary{}{\Wrappedafterbreak\char`\{}{\char`\{}}%
            \def\PYGZcb{\discretionary{\char`\}}{\Wrappedafterbreak}{\char`\}}}%
            \def\PYGZca{\discretionary{\char`\^}{\Wrappedafterbreak}{\char`\^}}%
            \def\PYGZam{\discretionary{\char`\&}{\Wrappedafterbreak}{\char`\&}}%
            \def\PYGZlt{\discretionary{}{\Wrappedafterbreak\char`\<}{\char`\<}}%
            \def\PYGZgt{\discretionary{\char`\>}{\Wrappedafterbreak}{\char`\>}}%
            \def\PYGZsh{\discretionary{}{\Wrappedafterbreak\char`\#}{\char`\#}}%
            \def\PYGZpc{\discretionary{}{\Wrappedafterbreak\char`\%}{\char`\%}}%
            \def\PYGZdl{\discretionary{}{\Wrappedafterbreak\char`\$}{\char`\$}}%
            \def\PYGZhy{\discretionary{\char`\-}{\Wrappedafterbreak}{\char`\-}}%
            \def\PYGZsq{\discretionary{}{\Wrappedafterbreak\textquotesingle}{\textquotesingle}}%
            \def\PYGZdq{\discretionary{}{\Wrappedafterbreak\char`\"}{\char`\"}}%
            \def\PYGZti{\discretionary{\char`\~}{\Wrappedafterbreak}{\char`\~}}%
        } 
        \newcommand*\Wrappedbreaksatpunct {%
            \lccode`\~`\.\lowercase{\def~}{\discretionary{\hbox{\char`\.}}{\Wrappedafterbreak}{\hbox{\char`\.}}}%
            \lccode`\~`\,\lowercase{\def~}{\discretionary{\hbox{\char`\,}}{\Wrappedafterbreak}{\hbox{\char`\,}}}%
            \lccode`\~`\;\lowercase{\def~}{\discretionary{\hbox{\char`\;}}{\Wrappedafterbreak}{\hbox{\char`\;}}}%
            \lccode`\~`\:\lowercase{\def~}{\discretionary{\hbox{\char`\:}}{\Wrappedafterbreak}{\hbox{\char`\:}}}%
            \lccode`\~`\?\lowercase{\def~}{\discretionary{\hbox{\char`\?}}{\Wrappedafterbreak}{\hbox{\char`\?}}}%
            \lccode`\~`\!\lowercase{\def~}{\discretionary{\hbox{\char`\!}}{\Wrappedafterbreak}{\hbox{\char`\!}}}%
            \lccode`\~`\/\lowercase{\def~}{\discretionary{\hbox{\char`\/}}{\Wrappedafterbreak}{\hbox{\char`\/}}}%
            \catcode`\.\active
            \catcode`\,\active 
            \catcode`\;\active
            \catcode`\:\active
            \catcode`\?\active
            \catcode`\!\active
            \catcode`\/\active 
            \lccode`\~`\~ 	
        }
    \makeatother

    \let\OriginalVerbatim=\Verbatim
    \makeatletter
    \renewcommand{\Verbatim}[1][1]{%
        \sbox\Wrappedcontinuationbox {\Wrappedcontinuationsymbol}%
        \sbox\Wrappedvisiblespacebox {\FV@SetupFont\Wrappedvisiblespace}%
        \def\FancyVerbFormatLine ##1{\hsize\linewidth
            \vtop{\raggedright\hyphenpenalty\z@\exhyphenpenalty\z@
                \doublehyphendemerits\z@\finalhyphendemerits\z@
                \strut ##1\strut}%
        }%
        \def\FV@Space {%
            \nobreak\hskip\z@ plus\fontdimen3\font minus\fontdimen4\font
            \discretionary{\copy\Wrappedvisiblespacebox}{\Wrappedafterbreak}
            {\kern\fontdimen2\font}%
        }%
        
        \Wrappedbreaksatspecials
        \OriginalVerbatim[#1,codes*=\Wrappedbreaksatpunct]%
    }
    \makeatother

    \definecolor{incolor}{HTML}{303F9F}
    \definecolor{outcolor}{HTML}{D84315}
    \definecolor{cellborder}{HTML}{CFCFCF}
    \definecolor{cellbackground}{HTML}{F7F7F7}
    
    \makeatletter
    \newcommand{\boxspacing}{\kern\kvtcb@left@rule\kern\kvtcb@boxsep}
    \makeatother
    \newcommand{\prompt}[4]{
        \ttfamily\llap{{\color{#2}[#3]:\hspace{3pt}#4}}\vspace{-\baselineskip}
    }

    \hypersetup{
      breaklinks=true,  
      colorlinks=true,
      urlcolor=urlcolor,
      linkcolor=linkcolor,
      citecolor=citecolor,
      }
    
\begin{document}
    
    \maketitle

    \hypertarget{using-bayesmixedlogit-and-bayesmixedlogitwtp-in-stata}{%
\section{\texorpdfstring{Introduction}{Using bayesmixedlogit and bayesmixedlogitwtp in Stata}}\label{using-bayesmixedlogit-and-bayesmixedlogitwtp-in-stata}}

This document presents an overview of the \texttt{bayesmixedlogit} and
\texttt{bayesmixedlogitwtp} \texttt{Stata} packages. It mirrors closely
the helpfile obtainable in \texttt{Stata} (i.e., through
\texttt{help\ bayesmixedlogit} or \texttt{help\ bayesmixedlogitwtp}).
Further background for the packages can be found in
\href{https://www.stata-journal.com/article.html?article=st0354}{Baker(2014)}.

    \hypertarget{description}{%
\subsection{Description}\label{description}}

\texttt{bayesmixedlogit} can be used to fit mixed logit models using
Bayesian methods -- more precisely, \texttt{bayesmixedlogit} produces
draws from the posterior parameter distribution and then presents
summary and other statistics describing the results of the drawing.
Detailed analysis of the draws is left to the discretion of the user.

Implementation of \texttt{bayesmixedlogit} follows
\href{https://eml.berkeley.edu/books/choice2.html}{Train (2009, chap.
12)}, and details of how the algorithm works are described in
\href{https://www.stata-journal.com/article.html?article=st0354}{Baker(2014)}.
A diffuse prior for the mean values of the random coefficients is
assumed, and the prior distribution on the covariance matrix of random
coefficients is taken to be an identity inverse Wishart.
\texttt{bayesmixedlogit} uses the \texttt{Mata} routines
\texttt{amcmc()} (if not installed: \texttt{ssc\ install\ amcmc}) for
adaptive Markov chain Monte Carlo sampling from the posterior
distribution of individual level coefficients and fixed coefficients.
The data setup for \texttt{bayesmixedlogit} is the same as for
\texttt{clogit} (\texttt{ssc\ install\ clogit}). Much of the syntax
follows that used by
\href{https://www.stata-journal.com/article.html?article=st0133}{Hole
(2007)} in development of the command \texttt{mixlogit}.

    \hypertarget{options}{%
\subsection{Options}\label{options}}

\textbf{\texttt{group}}(\emph{\texttt{varname}}) specifies a numeric
identifier variable for choice occasions. \textbf{\texttt{group}}() is
required.

\textbf{\texttt{identifier}}(\emph{\texttt{varname}}) identifies
coefficient sets (those observations for which a set of coefficients
apply). Thus, when a person is observed making choices over multiple
occasions, one would use
\textbf{\texttt{group}}(\emph{\texttt{varname}}) to specify the choice
occasions, while \textbf{\texttt{identifier}}(\emph{\texttt{varname}})
would identify the person. \textbf{\texttt{identifier}}() is required.

\textbf{\texttt{rand}}(\emph{\texttt{varlist}}) specifies independent
variables with random coefficients. The variables immediately following
the dependent variable in the syntax are considered to have fixed
coefficients (see the examples below). While a model can be run without
any independent variables with fixed coefficients, at least one
random-coefficient independent variable is required for bayesmixedlogit
to work. \textbf{\texttt{rand}}() is required.

\textbf{\texttt{draws}}(\texttt{\#}) specifies the number of draws that
are to be taken from the posterior distribution of the parameters. The
default is \textbf{\texttt{draws}}(\texttt{1000}).

\textbf{\texttt{drawsrandom}}(\#) is an advanced option. The drawing
algorithm treats each set of random coefficients as a Gibbs step in
sampling from the joint posterior distribution of parameters. In
difficult, large-dimensional problems, it might be desirable to let
individual Gibbs steps run for more than one draw to achieve better
mixing and convergence of the algorithm.

\textbf{\texttt{drawsfixed}}(\#) is a more advanced option. The drawing
algorithm treats fixed coefficients as a Gibbs step in sampling from the
joint posterior distribution of parameters. In difficult,
large-dimensional problems, it might be desirable to let this step in
Gibbs sampling run for more than a single draw. The default is
\textbf{\texttt{drawsfixed(1)}}.

\textbf{\texttt{burn}}(\#) specifies the length of the burn-in period;
the first \# draws are discarded upon completion of the algorithm and
before further results are computed.

\textbf{\texttt{thin}}(\#) specifies that only every \#th draw is to be
retained, so if thin(3) is specified, only every third draw is retained.
This option is designed to help ease autocorrelation in the resulting
draws, as is the option jumble, which randomly mixes draws. Both options
may be applied.

\textbf{\texttt{araterandom}}(\#) specifies the desired acceptance rate
for random coefficients and should be a number between zero and one.
Because an adaptive acceptance-rejection method is used to sample random
coefficients, by specifying the desired acceptance rate, the user has
some control over adaptation of the algorithm to the problem. The
default is \textbf{\texttt{araterandom}}(.234).

\textbf{\texttt{aratefixed}}(\#) specifies the desired acceptance rate
for fixed coefficients and works in the same fashion as
\textbf{\texttt{araterandom(\#)}}.

\textbf{\texttt{samplerrandom}}(\emph{\texttt{string}}) specifies the
type of sampler that is to be used when random parameters are drawn. It
may be set to either global or mwg. The default is
\textbf{\texttt{samplerrandom}}(\emph{\texttt{global}}), which means
that proposed changes to random parameters are drawn all at once. If
\texttt{mwg} -- an acronym for ``Metropolis within Gibbs'' -- is instead
chosen, each random parameter is drawn separately as an independent step
conditional on other random parameters in a nested Gibbs step.
\texttt{mwg} might be useful in situations in which initial values are
poorly scaled. The workings of these options are described in greater
detail in
\href{https://www.stata-journal.com/article.html?article=st0354}{Baker(2014)}.

\textbf{\texttt{samplerfixed}}(\emph{\texttt{string}}) specifies the
type of sampler that is used when fixed parameters are drawn. Options
are exactly as those described under
\textbf{\texttt{samplerrandom}}(\emph{\texttt{string}}).

\textbf{\texttt{dampparmfixed}}(\#) works exactly as option
\textbf{dampparmrandom}(\#) but is applied to drawing fixed parameters.

\textbf{\texttt{dampparmrandom}}(\#) is a parameter that controls how
aggressively the proposal distributions for random parameters are
adapted as drawing continues. If the parameter is set close to one,
adaptation is aggressive in its early phase of trying to achieve the
acceptance rate specified in \textbf{\texttt{araterandom}}(\#). If the
parameter is set closer to zero, adaptation is more gradual.

\textbf{\texttt{from}(\emph{\texttt{rowvector}}) specifies a row vector
of starting values for all parameters in order. If these are not
specified, starting values are obtained via estimation of a conditional
logit model via }\texttt{clogit}**.

\textbf{\texttt{fromvariance}}(\emph{\texttt{matrix}}) specifies a
matrix of starting values for the random parameters.

\textbf{\texttt{jumble}} specifies to randomly mix draws.

\textbf{\texttt{noisy}} specifies that a dot be produced every time a
complete pass through the algorithm is finished. After 50 iterations, a
function value \texttt{ln\_fc(p)} will be produced, which gives the
joint log of the value of the posterior choice probabilities evaluated
at the latest parameters. While \texttt{ln\_fc(p)} is not an objective
function per se, the author has found that drift in the value of this
function indicates that the algorithm has not yet converged or has other
problems.

\textbf{\texttt{saving}}(\emph{\texttt{filename}}) specifies a location
to store the draws from the distribution. The file will contain just the
draws after any burn-in period or thinning of values is applied.

\textbf{replace} specifies that an existing file is to be overwritten.

\textbf{append} specifies that an existing file is to be appended, which
might be useful if multiple runs need to be combined.

\textbf{\texttt{indsave}}(\emph{\texttt{filename}}) specifies a file to
which individual-level random parameters are to be saved. More
precisely, \textbf{\texttt{indsave}}(\emph{\texttt{filename}}) saves the
draws of the individual-level parameters. Caution: For long runs and
models with large numbers of individuals, specifying this option can
cause memory problems. Users should be careful how it is used and
consult some of the examples before employing the option.

\textbf{\texttt{indkeep}}(\#) is for use with indsave and specifies that
only the last \# draws of the individual-level random parameters be
kept. This helps avoid excessive memory consumption.

\textbf{\texttt{indwide}} is for use with indsave and affords the user a
degree of control over how individual-level parameters are saved. By
default, individual-level parameters are saved in a panel form, meaning
that each random parameter draw is saved in a row, where draws are
marked by the group identifier. If instead the user would prefer that
each row contain draws of each parameter, one could specify the indwide
option, which saves all draws in a single row, with the first entry of
the row being the group identifier. By analogy with \texttt{reshape}, by
default draws are saved in ``long'' format, whereas
\textbf{\texttt{indwide}} stores the draws in ``wide'' format.

\textbf{\texttt{replaceind}} functions in the same way as replace, but
in reference to the file specified in \textbf{\texttt{indsave}}.

\textbf{\texttt{appendind}} functions in the same way as
\textbf{\texttt{append}}, but in reference to the file specified in
\textbf{\texttt{indsave}}.

    \hypertarget{examples}{%
\subsection{Examples}\label{examples}}

    \hypertarget{example-1}{%
\subsubsection{Example 1}\label{example-1}}

A single random coefficient, one decision per group. The random
parameter acceptance rate is set to 0.4, and a total of 4,000 draws are
taken. The first 1,000 draws are dropped, and then every fifth draw is
retained. Draws are saved as \texttt{choice\_draws.dta}:

    \begin{tcolorbox}[breakable, size=fbox, boxrule=1pt, pad at break*=1mm,colback=cellbackground, colframe=cellborder]
\prompt{In}{incolor}{1}{\boxspacing}
\begin{Verbatim}[commandchars=\\\{\}]
\PY{k}{webuse} choice
\PY{k}{describe}
\end{Verbatim}
\end{tcolorbox}

    \begin{Verbatim}[commandchars=\\\{\}]



Contains data from http://www.stata-press.com/data/r14/choice.dta
  obs:           885
 vars:             7                          2 Dec 2014 13:25
 size:         9,735
--------------------------------------------------------------------------------
              storage   display    value
variable name   type    format     label      variable label
--------------------------------------------------------------------------------
id              int     \%9.0g
sex             byte    \%9.0g      sex
income          float   \%9.0g                 income in thousands
car             byte    \%9.0g      nation     nationality of car
size            byte    \%9.0g
choice          byte    \%9.0g                 ID's chosen car
dealer          byte    \%9.0g                 number of dealers of each
                                                nationality in ID's city
--------------------------------------------------------------------------------
Sorted by: id
    \end{Verbatim}

    \begin{tcolorbox}[breakable, size=fbox, boxrule=1pt, pad at break*=1mm,colback=cellbackground, colframe=cellborder]
\prompt{In}{incolor}{2}{\boxspacing}
\begin{Verbatim}[commandchars=\\\{\}]
bayesmixedlogit choice, rand(dealer) \PY{n+nf}{group}(id) id(id) \PY{c+cs}{///}
     draws(\PY{l+m}{4000}) burn(\PY{l+m}{1000}) thin(\PY{l+m}{5}) arater(.\PY{l+m}{4}) saving(choice\PYZus{}draws)\PY{k}{ replace}
\end{Verbatim}
\end{tcolorbox}

    \begin{Verbatim}[commandchars=\\\{\}]


Bayesian Mixed Logit Model                         Observations    =       885
                                                   Groups          =       295
Acceptance rates:                                  Choices         =       295
 Fixed coefs              =                        Total draws     =      4000
 Random coefs(ave,min,max)= 0.239, 0.186, 0.285    Burn-in draws   =      1000
                                                   *One of every 5 draws kept
------------------------------------------------------------------------------
      choice |      Coef.   Std. Err.      t    P>|t|     [95\% Conf. Interval]
-------------+----------------------------------------------------------------
Random       |
      dealer |   .2150072   .0360706     5.96   0.000     .1441673    .2858471
-------------+----------------------------------------------------------------
Cov\_Random   |
  var\_dealer |   .1024538   .0322976     3.17   0.002     .0390238    .1658839
------------------------------------------------------------------------------
   Draws saved in choice\_draws.dta


   Attention!
   *Results are presented to conform with Stata covention, but
    are summary statistics of draws, not coefficient estimates.
    \end{Verbatim}

    \hypertarget{example-2}{%
\subsubsection{Example 2}\label{example-2}}

Fitting a mixed logit model using \texttt{bayesmixedlogit}, using the
methods as described in
\href{https://www.stata-press.com/books/regression-models-categorical-dependent-variables/long3-brochure.pdf}{Long
and Freese (2006, sec.~7.2.4)}. The data must first be rendered into the
correct format, which can be done using the command \texttt{case2alt},
which is part of the package \texttt{spost9\_ado}; if not installed,
type
\texttt{net\ install\ spost9\_ado,\ from(https://jslsoc.sitehost.iu.edu/stata)}
from the \texttt{Stata} prompt. The example first arranges the data and
then generates and summarizes posterior draws from a mixed logit model.
The model uses \texttt{bangladesh.dta}, which has information on
contraceptive choice by a series of families. Coefficients of
explanatory variables vary at the district level.

    \begin{tcolorbox}[breakable, size=fbox, boxrule=1pt, pad at break*=1mm,colback=cellbackground, colframe=cellborder]
\prompt{In}{incolor}{3}{\boxspacing}
\begin{Verbatim}[commandchars=\\\{\}]
\PY{k}{webuse} choice
\PY{k}{describe}
\end{Verbatim}
\end{tcolorbox}

    \begin{Verbatim}[commandchars=\\\{\}]



Contains data from http://www.stata-press.com/data/r14/choice.dta
  obs:           885
 vars:             7                          2 Dec 2014 13:25
 size:         9,735
--------------------------------------------------------------------------------
              storage   display    value
variable name   type    format     label      variable label
--------------------------------------------------------------------------------
id              int     \%9.0g
sex             byte    \%9.0g      sex
income          float   \%9.0g                 income in thousands
car             byte    \%9.0g      nation     nationality of car
size            byte    \%9.0g
choice          byte    \%9.0g                 ID's chosen car
dealer          byte    \%9.0g                 number of dealers of each
                                                nationality in ID's city
--------------------------------------------------------------------------------
Sorted by: id
    \end{Verbatim}

    \begin{tcolorbox}[breakable, size=fbox, boxrule=1pt, pad at break*=1mm,colback=cellbackground, colframe=cellborder]
\prompt{In}{incolor}{4}{\boxspacing}
\begin{Verbatim}[commandchars=\\\{\}]
\PY{k}{webuse} bangladesh,\PY{k}{ clear}
case2alt, casevars(urban age) choice(c\PYZus{}use)\PY{k}{ gen}(choice)
\end{Verbatim}
\end{tcolorbox}

    \begin{Verbatim}[commandchars=\\\{\}]

(Bangladesh Fertility Survey, 1989)

(note: variable \_id used since case() not specified)
(note: variable \_altnum used since altnum() not specified)

choice indicated by: choice
case identifier: \_id
case-specific interactions: no* yes*
    \end{Verbatim}

    \begin{tcolorbox}[breakable, size=fbox, boxrule=1pt, pad at break*=1mm,colback=cellbackground, colframe=cellborder]
\prompt{In}{incolor}{5}{\boxspacing}
\begin{Verbatim}[commandchars=\\\{\}]
bayesmixedlogit choice, rand(yesXurban yesXage yes) \PY{n+nf}{group}(\PYZus{}id) id(district) \PY{c+cs}{///}
     draws(\PY{l+m}{10000}) burn(\PY{l+m}{5000}) saving(bdesh\PYZus{}draws)\PY{k}{ replace}
\end{Verbatim}
\end{tcolorbox}

    \begin{Verbatim}[commandchars=\\\{\}]


Bayesian Mixed Logit Model                         Observations    =      3868
                                                   Groups          =        60
Acceptance rates:                                  Choices         =      1934
 Fixed coefs              =                        Total draws     =     10000
 Random coefs(ave,min,max)= 0.310, 0.231, 0.354    Burn-in draws   =      5000
-------------------------------------------------------------------------------
       choice |      Coef.   Std. Err.      t    P>|t|     [95\% Conf. Interval]
--------------+----------------------------------------------------------------
Random        |
    yesXurban |   .7796584   .1905085     4.09   0.000     .4061781    1.153139
      yesXage |   .0067593   .0329065     0.21   0.837    -.0577519    .0712706
          yes |  -.7777077   .1154274    -6.74   0.000    -1.003996   -.5514193
--------------+----------------------------------------------------------------
Cov\_Random    |
var\_yesXurban |   .9302604   .4088646     2.28   0.023     .1287065    1.731814
cov\_yesXurb\textasciitilde{}e |  -.0010909    .034572    -0.03   0.975    -.0688672    .0666853
cov\_yesXurb\textasciitilde{}s |  -.3946547    .225488    -1.75   0.080    -.8367101    .0474007
  var\_yesXage |   .0611443   .0120586     5.07   0.000     .0375042    .0847844
cov\_yesXage\textasciitilde{}s |   .0051727   .0258404     0.20   0.841    -.0454859    .0558313
      var\_yes |   .5352136   .1687242     3.17   0.002     .2044401    .8659871
-------------------------------------------------------------------------------
   Draws saved in bdesh\_draws.dta


   Attention!
   *Results are presented to conform with Stata covention, but
    are summary statistics of draws, not coefficient estimates.
    \end{Verbatim}

    Suppose one wished to save some values of individual-level random
parameters, but that the problem has too many individuals or requires
too many draws to get to convergence. A useful approach in these
circumstances is to complete a long first run without saving parameters,
and then do a short second one using starting values. Suppose that the
code in the previous example has been run. One can then run something to
the effect of the following to get individual parameters:

    \begin{tcolorbox}[breakable, size=fbox, boxrule=1pt, pad at break*=1mm,colback=cellbackground, colframe=cellborder]
\prompt{In}{incolor}{6}{\boxspacing}
\begin{Verbatim}[commandchars=\\\{\}]
\PY{k}{mat} b =\PY{k}{ e}(b)
\PY{k}{mat} beta = b[\PY{l+m}{1}, \PY{l+m}{1}..\PY{l+m}{3}]
\PY{k}{mat} V = b[\PY{l+m}{1},\PY{l+m}{4}], b[\PY{l+m}{1},\PY{l+m}{5}], b[\PY{l+m}{1},\PY{l+m}{6}] \PYZbs{} b[\PY{l+m}{1},\PY{l+m}{5}], b[\PY{l+m}{1},\PY{l+m}{7}], b[\PY{l+m}{1},\PY{l+m}{8}] \PYZbs{} b[\PY{l+m}{1},\PY{l+m}{6}], b[\PY{l+m}{1},\PY{l+m}{7}], b[\PY{l+m}{1},\PY{l+m}{9}]
\end{Verbatim}
\end{tcolorbox}

    \begin{tcolorbox}[breakable, size=fbox, boxrule=1pt, pad at break*=1mm,colback=cellbackground, colframe=cellborder]
\prompt{In}{incolor}{7}{\boxspacing}
\begin{Verbatim}[commandchars=\\\{\}]
bayesmixedlogit choice, rand(yesXurban yesXage yes) \PY{n+nf}{group}(\PYZus{}id) id(district) \PY{c+cs}{///}
     from(beta) fromv(V) draws(\PY{l+m}{100}) indsave(randpars) indkeep(\PY{l+m}{50}) replaceind
\end{Verbatim}
\end{tcolorbox}

    \begin{Verbatim}[commandchars=\\\{\}]


Bayesian Mixed Logit Model                         Observations    =      3868
                                                   Groups          =        60
Acceptance rates:                                  Choices         =      1934
 Fixed coefs              =                        Total draws     =       100
 Random coefs(ave,min,max)= 0.297, 0.160, 0.400    Burn-in draws   =         0
-------------------------------------------------------------------------------
       choice |      Coef.   Std. Err.      t    P>|t|     [95\% Conf. Interval]
--------------+----------------------------------------------------------------
Random        |
    yesXurban |   .7763147   .1041361     7.45   0.000     .5697116    .9829178
      yesXage |   .0078787   .0344768     0.23   0.820    -.0605223    .0762797
          yes |  -.7578993   .1074017    -7.06   0.000    -.9709811   -.5448174
--------------+----------------------------------------------------------------
Cov\_Random    |
var\_yesXurban |   .3507105   .2150101     1.63   0.106    -.0758635    .7772845
cov\_yesXurb\textasciitilde{}e |  -.0014138   .0203943    -0.07   0.945    -.0418756    .0390479
cov\_yesXurb\textasciitilde{}s |  -.1896818    .140957    -1.35   0.181    -.4693365    .0899729
  var\_yesXage |   .0599069   .0127719     4.69   0.000     .0345679    .0852459
cov\_yesXage\textasciitilde{}s |   .0070305   .0200905     0.35   0.727    -.0328285    .0468896
      var\_yes |   .3150498   .1372705     2.30   0.024      .042709    .5873906
-------------------------------------------------------------------------------
   50 value(s) of individual-level random parameters saved in randpars.dta


   Attention!
   *Results are presented to conform with Stata covention, but
    are summary statistics of draws, not coefficient estimates.
    \end{Verbatim}

    One post-estimation idea is to get the mean for parameter values by
individuals, and fit some kernel density to the means to view the
distribution of the individual-level parameters:

    \begin{tcolorbox}[breakable, size=fbox, boxrule=1pt, pad at break*=1mm,colback=cellbackground, colframe=cellborder]
\prompt{In}{incolor}{8}{\boxspacing}
\begin{Verbatim}[commandchars=\\\{\}]
\PY{k}{bysort} district:\PY{k}{ egen} myesXurban =\PY{k}{ mean}(yesXurban)
\PY{k}{bysort} district:\PY{k}{ gen} last = \PYZus{}n \PY{o}{==} \PYZus{}N
\PY{k}{kdensity} myesXurban\PY{k}{ if} last
\PY{k}{graph}\PY{k}{ display}
\end{Verbatim}
\end{tcolorbox}

    \begin{center}
    \adjustimage{max size={0.9\linewidth}{0.9\paperheight}}{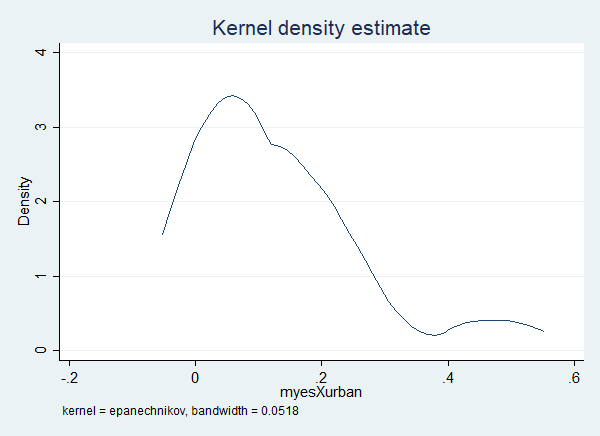}
    \end{center}
    { \hspace*{\fill} \\}
    
    \begin{Verbatim}[commandchars=\\\{\}]




.     global stata\_kernel\_graph\_counter = \$stata\_kernel\_graph\_counter + 1
    \end{Verbatim}

    \hypertarget{example-3}{%
\subsubsection{Example 3}\label{example-3}}

Of course, it is possible to just do one run and retain all the
information. As one final example:

    \begin{tcolorbox}[breakable, size=fbox, boxrule=1pt, pad at break*=1mm,colback=cellbackground, colframe=cellborder]
\prompt{In}{incolor}{9}{\boxspacing}
\begin{Verbatim}[commandchars=\\\{\}]
\PY{k}{webuse} union,\PY{k}{ clear}
case2alt, casevars(age grade) choice(union)\PY{k}{ gen}(unionmember)
bayesmixedlogit unionmember, rand(y0Xage y0Xgrade y0) \PY{n+nf}{group}(\PYZus{}id) id(idcode) \PY{c+cs}{///}
     draws(\PY{l+m}{1000}) burn(\PY{l+m}{800}) saving(parm\PYZus{}draws) indsave(indparm\PYZus{}draws) indkeep(\PY{l+m}{20}) replaceind\PY{k}{ replace}
\end{Verbatim}
\end{tcolorbox}

    \begin{Verbatim}[commandchars=\\\{\}]

(NLS Women 14-24 in 1968)

(note: variable \_id used since case() not specified)
(note: variable \_altnum used since altnum() not specified)

choice indicated by: unionmember
case identifier: \_id
case-specific interactions: y0* y1*



Bayesian Mixed Logit Model                         Observations    =     52400
                                                   Groups          =      4434
Acceptance rates:                                  Choices         =     26200
 Fixed coefs              =                        Total draws     =      1000
 Random coefs(ave,min,max)= 0.221, 0.017, 0.339    Burn-in draws   =       800
-------------------------------------------------------------------------------
  unionmember |      Coef.   Std. Err.      t    P>|t|     [95\% Conf. Interval]
--------------+----------------------------------------------------------------
Random        |
       y0Xage |   .0427559   .0045651     9.37   0.000      .033754    .0517578
     y0Xgrade |  -.0475095   .0097271    -4.88   0.000    -.0666903   -.0283288
           y0 |   2.405118   .0067917   354.13   0.000     2.391726    2.418511
--------------+----------------------------------------------------------------
Cov\_Random    |
   var\_y0Xage |   .0472717   .0030153    15.68   0.000     .0413258    .0532176
cov\_y0Xagey\textasciitilde{}e |   -.084886   .0067274   -12.62   0.000    -.0981517   -.0716203
 cov\_y0Xagey0 |  -.0017989   .0019461    -0.92   0.356    -.0056365    .0020387
 var\_y0Xgrade |   .2070458   .0164691    12.57   0.000     .1745704    .2395211
cov\_y0Xgrad\textasciitilde{}0 |  -.0031688   .0034891    -0.91   0.365     -.010049    .0037114
       var\_y0 |   .0808648   .0067269    12.02   0.000     .0676001    .0941296
-------------------------------------------------------------------------------
   Draws saved in parm\_draws.dta
   20 value(s) of individual-level random parameters saved in indparm\_draws.dta


   Attention!
   *Results are presented to conform with Stata covention, but
    are summary statistics of draws, not coefficient estimates.
    \end{Verbatim}

    \hypertarget{stored-results}{%
\subsubsection{Stored results}\label{stored-results}}

\textbf{\texttt{bayesmixedlogit}} stores the following in e():

\hypertarget{scalars}{%
\paragraph{Scalars}\label{scalars}}

\textbf{\texttt{e}}(\texttt{N}) number of observations

\textbf{\texttt{e}}(\texttt{df\_r}) degrees of freedom for summarizing
draws (equal to number of retained draws)

\textbf{\texttt{e}}(\texttt{krnd}) number of random parameters

\textbf{\texttt{e}}(\texttt{kfix}) number of fixed parameters

\textbf{\texttt{e}}(\texttt{draws}) number of draws

\textbf{\texttt{e}}(\texttt{burn}) burn-in observations

\textbf{\texttt{e}}(\texttt{thin}) thinning parameter

\textbf{\texttt{e}}(\texttt{random\_draws}) number of draws of each set
of random parameters per pass

\textbf{\texttt{e}}(\texttt{fixed\_draws}) number of draws of fixed
parameters per pass

\textbf{\texttt{e}}(\texttt{damper\_fixed}) damping parameter -- fixed
parameters

\textbf{\texttt{e}}(\texttt{damper\_random}) damping parameter -- random
parameters

\textbf{\texttt{e}}(\texttt{opt\_arate\_fixed}) desired acceptance rate
-- fixed parameters

\textbf{\texttt{e}}(\texttt{opt\_arate\_random}) desired acceptance rate
-- random parameters

\textbf{\texttt{e}}(\texttt{N\_groups}) number of groups

\textbf{\texttt{e}}(\texttt{N\_choices}) number of choice occasions

\textbf{\texttt{e}}(\texttt{arates\_fa}) acceptance rate -- fixed
parameters

\textbf{\texttt{e}}(\texttt{arates\_ra}) average acceptance rate --
random parameters

\textbf{\texttt{e}}(\texttt{arates\_rmax}) maximum acceptance rate --
random parameters

\textbf{\texttt{e}}(\texttt{arates\_rmin}) minimum acceptance rate --
random parameters

\textbf{\texttt{e}}(\texttt{inddraws}) draws of individual parameters
kept

\hypertarget{macros}{%
\paragraph{Macros}\label{macros}}

\textbf{\texttt{e}}(\texttt{cmd}) \textbf{\texttt{bayesmixedlogit}}

\textbf{\texttt{e}}(\texttt{depvar}) name of dependent variable

\textbf{\texttt{e}}(\texttt{indepvars}) independent variables

\textbf{\texttt{e}}(\texttt{title}) title in estimation output

\textbf{\texttt{e}}(\texttt{properties}) b V

\textbf{\texttt{e}}(\texttt{saving}) file containing results

\textbf{\texttt{e}}(\texttt{fixed\_sampler}) sampler type for fixed
parameters

\textbf{\texttt{e}}(\texttt{random\_sampler}) sampler type for random
parameters

\textbf{\texttt{e}}(\texttt{random}) random parameter names

\textbf{\texttt{e}}(\texttt{fixed}) fixed parameter names

\textbf{\texttt{e}}(\texttt{identifier}) identifier for individuals

\textbf{\texttt{e}}(\texttt{group}) identifier for choice occasions

\textbf{\texttt{e}}(\texttt{indsave}) file holding individual-level
parameter draws

\hypertarget{matrices}{%
\paragraph{Matrices}\label{matrices}}

\textbf{\texttt{e}}(\texttt{b}) mean parameter values

\textbf{\texttt{e}}(\texttt{V}) variance-covariance matrix of parameters

\textbf{\texttt{e}}(\texttt{V\_init}) initial variance-covariance matrix
of random parameters

\textbf{\texttt{e}}(\texttt{b\_init}) initial mean vector of random
parameters

\textbf{\texttt{e}}(\texttt{arates\_fixed}) row vector of acceptance
rates of fixed parameters

\textbf{\texttt{e}}(\texttt{arates\_rand}) vector or matrix of
acceptance rates of random parameters

\hypertarget{functions}{%
\paragraph{Functions}\label{functions}}

\textbf{\texttt{e}}(\texttt{sample}) marks estimation sample

    \hypertarget{bayesmixedlogitwtp}{%
\subsection{\texorpdfstring{\texttt{bayesmixedlogitwtp}}{bayesmixedlogitwtp}}\label{bayesmixedlogitwtp}}

\texttt{bayesmixedlogitwtp} is essentially a wrapper for
\texttt{bayesmixedlogit}, with a transformation of the coefficient on a
price variable. The defining characteristic of the WTP-space mixed logit
model is normalization of coefficients using the (random) coefficient on
a designated price variable, as described in
\href{https://link.springer.com/chapter/10.1007/1-4020-3684-1_1}{Train
and Weeks (2005)},
\href{https://econpapers.repec.org/article/oupajagec/v_3a90_3ay_3a2008_3ai_3a4_3ap_3a994-1010.htm}{Scarpa,
Thiene, and Train (2008)}, and
\href{https://link.springer.com/article/10.1007/s00181-011-0500-1}{Hole
and Kolstad (2012)}.

The model assumes that the coefficient on the price variable follows
(the negative of) a log-normal distribution. Hence, if the estimated
parameter is \textbf{\texttt{b}}, the price variable has coefficient
\textbf{\texttt{-exp(b)}}. The transformed coefficient is saved and
displayed as part of the output, but as presented the saved and display
value is the negative of the exponentiated average value of b, not the
average of the value \textbf{\texttt{-exp(b)}}.

\hypertarget{options}{%
\subsection{Options}\label{options}}

All options for \texttt{bayesmixedlogitwtp} are the same as
\texttt{bayesmixedlogit}, with the following additional option:

\textbf{\texttt{price}}(\emph{\texttt{varname}}) specifies a numeric
identifier variable for price occasions. \textbf{\texttt{price}}() is
required.

\hypertarget{stored-results}{%
\subsubsection{Stored results}\label{stored-results}}

In addition to all the scalars, macros, and matrices stored by
\texttt{bayesmixedlogit}, \texttt{bayesmixedlogitwtp} adds the following
additional macros:

\hypertarget{scalars}{%
\paragraph{Scalars}\label{scalars}}

\texttt{e(price\_coef)} - exponent of mean of estimated coefficient on
price variable

\hypertarget{macros}{%
\paragraph{Macros}\label{macros}}

\texttt{e(pricevar)} - name of price variable

\#\#\# Example 4

    The following example mirrors examples provided of usage of the
mixlogitwtp command (with thanks to Arne Rise Hole for allowing use of
the example):

    \begin{tcolorbox}[breakable, size=fbox, boxrule=1pt, pad at break*=1mm,colback=cellbackground, colframe=cellborder]
\prompt{In}{incolor}{10}{\boxspacing}
\begin{Verbatim}[commandchars=\\\{\}]
\PY{k}{use} http:\PY{o}{/}\PY{o}{/}fmwww\PY{l+m}{.}bc\PY{l+m}{.}edu\PY{o}{/}repec\PY{o}{/}bocode\PY{o}{/}t\PY{o}{/}traindata\PY{l+m}{.}dta,\PY{k}{ clear}
\PY{k}{describe}

bayesmixedlogitwtp y\PY{k}{ contract}\PY{k}{ local} wknown, \PY{n+nf}{group}(gid) id(pid) price(price) \PY{c+cs}{///}
     rand(seasonal tod) draws(\PY{l+m}{4000}) burn(\PY{l+m}{1000}) thin(\PY{l+m}{5}) arater(.\PY{l+m}{4}) saving(draws)\PY{k}{ replace}
\end{Verbatim}
\end{tcolorbox}

    \begin{Verbatim}[commandchars=\\\{\}]



Contains data from http://fmwww.bc.edu/repec/bocode/t/traindata.dta
  obs:         4,780
 vars:             9                          28 Nov 2006 18:40
 size:        52,580
--------------------------------------------------------------------------------
              storage   display    value
variable name   type    format     label      variable label
--------------------------------------------------------------------------------
y               byte    \%8.0g
price           byte    \%8.0g
contract        byte    \%8.0g
local           byte    \%8.0g
wknown          byte    \%8.0g
tod             byte    \%8.0g
seasonal        byte    \%8.0g
gid             int     \%8.0g
pid             int     \%9.0g
--------------------------------------------------------------------------------
Sorted by:



Bayesian Mixed Logit Model - WTP Form              Observations    =      4780
                                                   Groups          =       100
Acceptance rates:                                  Choices         =      1195
 Fixed coefs              = 0.290                  Total draws     =      4000
 Random coefs(ave,min,max)= 0.214, 0.164, 0.260    Burn-in draws   =      1000
                                                   *One of every 5 draws kept
-------------------------------------------------------------------------------
            y |      Coef.   Std. Err.      t    P>|t|     [95\% Conf. Interval]
--------------+----------------------------------------------------------------
Fixed         |
     contract |   -.249242   .0298574    -8.35   0.000    -.3078797   -.1906042
        local |   2.425951   .2125614    11.41   0.000     2.008497    2.843406
       wknown |   1.741476   .1599273    10.89   0.000     1.427391    2.055562
--------------+----------------------------------------------------------------
Random        |
        price |  -.3333228   .0947416    -3.52   0.000    -.5193883   -.1472574
     seasonal |  -9.779766   .3131075   -31.23   0.000    -10.39469   -9.164847
          tod |  -9.612307   .3512882   -27.36   0.000    -10.30221   -8.922403
--------------+----------------------------------------------------------------
Cov\_Random    |
    var\_price |   .2942746   .0795158     3.70   0.000     .1381115    .4504377
cov\_pricese\textasciitilde{}l |  -.2675019   .2485057    -1.08   0.282    -.7555485    .2205448
 cov\_pricetod |  -.4965004   .2658258    -1.87   0.062    -1.018562    .0255617
 var\_seasonal |   5.859335   1.728996     3.39   0.001     2.463715    9.254955
cov\_seasona\textasciitilde{}d |   4.173949   1.390441     3.00   0.003     1.443226    6.904672
      var\_tod |   7.391194   1.994975     3.70   0.000     3.473211    11.30918
-------------------------------------------------------------------------------
   Draws saved in draws.dta
The price variable is price with transformed coef (-exp(b)): -0.779


   Attention!
   *Results are presented to conform with Stata covention, but
    are summary statistics of draws, not coefficient estimates.
    \end{Verbatim}

    \hypertarget{example-5}{%
\subsubsection{Example 5}\label{example-5}}

A case in which all coefficients are random:

    \begin{tcolorbox}[breakable, size=fbox, boxrule=1pt, pad at break*=1mm,colback=cellbackground, colframe=cellborder]
\prompt{In}{incolor}{11}{\boxspacing}
\begin{Verbatim}[commandchars=\\\{\}]
bayesmixedlogitwtp y, \PY{n+nf}{group}(gid) id(pid) price(price) rand(seasonal tod wknown) \PY{c+cs}{///}
     draws(\PY{l+m}{2000}) burn(\PY{l+m}{1000}) thin(\PY{l+m}{5}) arater(.\PY{l+m}{4}) saving(draws)\PY{k}{ replace}
\end{Verbatim}
\end{tcolorbox}

    \begin{Verbatim}[commandchars=\\\{\}]


Bayesian Mixed Logit Model - WTP Form              Observations    =      4780
                                                   Groups          =       100
Acceptance rates:                                  Choices         =      1195
 Fixed coefs              =                        Total draws     =      2000
 Random coefs(ave,min,max)= 0.152, 0.068, 0.237    Burn-in draws   =      1000
                                                   *One of every 5 draws kept
-------------------------------------------------------------------------------
            y |      Coef.   Std. Err.      t    P>|t|     [95\% Conf. Interval]
--------------+----------------------------------------------------------------
Random        |
        price |  -.6533948   .1186226    -5.51   0.000    -.8873062   -.4194833
     seasonal |  -9.684424   .4084787   -23.71   0.000     -10.4899   -8.878947
          tod |  -9.844531   .4323386   -22.77   0.000    -10.69706   -8.992004
       wknown |   .9696596   .2220143     4.37   0.000     .5318704    1.407449
--------------+----------------------------------------------------------------
Cov\_Random    |
    var\_price |   .7395054   .2301116     3.21   0.002     .2857491    1.193262
cov\_pricese\textasciitilde{}l |  -.4428133   .4570225    -0.97   0.334    -1.344014    .4583876
 cov\_pricetod |  -.4947435   .5077538    -0.97   0.331    -1.495981    .5064943
cov\_pricewk\textasciitilde{}n |  -.2005007   .2009999    -1.00   0.320    -.5968516    .1958502
 var\_seasonal |   7.062787   2.070746     3.41   0.001     2.979491    11.14608
cov\_seasona\textasciitilde{}d |   5.437463   1.669412     3.26   0.001     2.145556    8.729371
cov\_seasona\textasciitilde{}n |  -.8402783   .5547666    -1.51   0.131     -1.93422    .2536639
      var\_tod |   9.310019   2.347582     3.97   0.000      4.68083    13.93921
cov\_todwknown |  -1.204763   .5393129    -2.23   0.027    -2.268232   -.1412935
   var\_wknown |   .9792376   .2518897     3.89   0.000     .4825372    1.475938
-------------------------------------------------------------------------------
   Draws saved in draws.dta
The price variable is price with transformed coef (-exp(b)): -0.520


   Attention!
   *Results are presented to conform with Stata covention, but
    are summary statistics of draws, not coefficient estimates.
    \end{Verbatim}

    Looking at the distribution of draws for the \texttt{price} variable:

    \begin{tcolorbox}[breakable, size=fbox, boxrule=1pt, pad at break*=1mm,colback=cellbackground, colframe=cellborder]
\prompt{In}{incolor}{12}{\boxspacing}
\begin{Verbatim}[commandchars=\\\{\}]
\PY{k}{use} draws,\PY{k}{ clear}
\PY{k}{describe}
\PY{k}{hist} price
\PY{k}{graph}\PY{k}{ display}
\end{Verbatim}
\end{tcolorbox}

    \begin{Verbatim}[commandchars=\\\{\}]



Contains data from draws.dta
  obs:           200
 vars:            16                          1 Mar 2021 08:53
 size:        12,800
--------------------------------------------------------------------------------
              storage   display    value
variable name   type    format     label      variable label
--------------------------------------------------------------------------------
price           float   \%10.0g
seasonal        float   \%10.0g
tod             float   \%10.0g
wknown          float   \%10.0g
var\_price       float   \%10.0g
cov\_priceseas\textasciitilde{}l float   \%10.0g
cov\_pricetod    float   \%10.0g
cov\_pricewknown float   \%10.0g
var\_seasonal    float   \%10.0g
cov\_seasonaltod float   \%10.0g
cov\_seasonalw\textasciitilde{}n float   \%10.0g
var\_tod         float   \%10.0g
cov\_todwknown   float   \%10.0g
var\_wknown      float   \%10.0g
fun\_val         float   \%10.0g
t               float   \%9.0g
--------------------------------------------------------------------------------
Sorted by:

(bin=14, start=-.98271137, width=.05126763)
    \end{Verbatim}

    \begin{center}
    \adjustimage{max size={0.9\linewidth}{0.9\paperheight}}{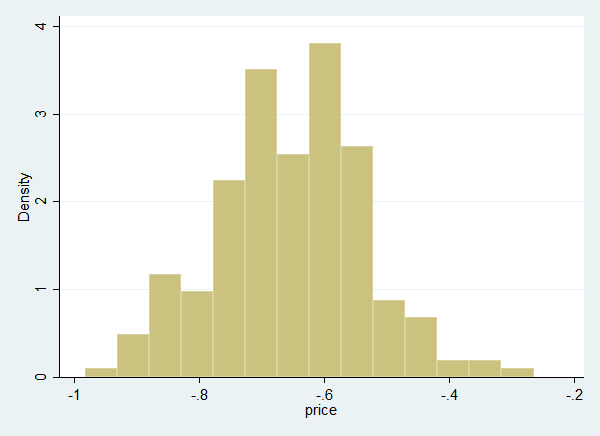}
    \end{center}
    { \hspace*{\fill} \\}
    
    \begin{Verbatim}[commandchars=\\\{\}]

.     global stata\_kernel\_graph\_counter = \$stata\_kernel\_graph\_counter + 1
    \end{Verbatim}

    \hypertarget{references}{%
\subsection{References}\label{references}}

Baker, M. J. 2014. \emph{Adaptive Markov chain Monte Carlo sampling and
estimation in \texttt{Mata}}. \textbf{Stata Journal} 14: 623-61.

Gelman, A., J. B. Carlin, H. S. Stern, and D. B. Rubin. 2009.
\textbf{Bayesian data analysis}. 2nd. ed. Boca Raton, FL: Chapman \&
Hall/CRC.

Hole, A. R. 2007. \emph{Fitting mixed logit models by using maximum
simulated likelihood}. \textbf{Stata Journal} 7: 388-401.

Hole, A. R. and J. R. Kolstad. 2012. \emph{Mixed logit estimation of
willingness to pay distributions: a comparison of models in preference
and WTP space using data from a health-related choice experiment}.
\textbf{Empirical Economics} 42: 445-469.

Long, J. S., and J. Freese. 2006. \textbf{Regression Models for
Categorical Dependent Variables Using Stata}. 2nd ed. College Station,
TX: Stata Press.

R. Scarpa, M. Thiene, and K. Train. 2008. \emph{Utility in willingness
to pay space: A tool to address confounding random scale effects in
destination choice to the Alps}. \textbf{American Journal of
Agricultural Economics} 90: 994-1010.

Train, K. E. 2009. \textbf{Discrete Choice Methods with Simulation}. 2nd
ed. Cambridge: Cambridge University Press.

Train, K. E. and M. Weeks. 2005. \emph{Discrete choice models in
preference space and willingness-to-pay space}. In: Scarpa R, Alberini A
(eds), \textbf{Application of simulation methods in environmental and
resource economics}. Springer, Dordrecht, pp 1-16.


\end{document}